\documentclass[twocolumn]{aastex6}

\begin{document}
\title{Evidence of Environmental Quenching at Redshift $z\approx 2$}
\author{Zhiyuan Ji\altaffilmark{1},
Mauro Giavalisco\altaffilmark{1},
Christina C. Williams\altaffilmark{2},
Sandra M. Faber\altaffilmark{3},
Henry C. Ferguson\altaffilmark{4},
Yicheng Guo\altaffilmark{3,7},
Teng Liu \altaffilmark{1,6},
Bomee Lee \altaffilmark{1,5}
}
\altaffiltext{1}{University of Massachusetts Amherst, 710 North Pleasant St, Amherst, MA 01003-9305, USA, zhiyuanji@astro.umass.edu}
\altaffiltext{2}{Steward Observatory, 933 N. Cherry Ave., University of Arizona, Tucson, AZ 85721, USA}
\altaffiltext{3}{University of California Observatories/Lick Observatory, University of California, Santa Cruz, CA 95064, USA}
\altaffiltext{4}{Space Telescope Science Institute, 3700 San Martin Boulevard, Baltimore, MD, 21218, USA}
\altaffiltext{5}{Infrared Processing and Analysis Center, California Institute of Technology, Pasadena, CA 91125, USA}
\altaffiltext{6}{University of Science and Technology of China, Hefei, Anhui, 230026, China}
\altaffiltext{7}{Department of Physics and Astronomy, University of Missouri, Columbia, MO, 65211, USA}

\begin{abstract}
We report evidence of ``environmental quenching'' among galaxies at redshift
$\rm{\approx2}$, namely the probability that a galaxy quenches its star
formation activity is enhanced in the regions of space in proximity of other
quenched, more massive galaxies. The effect is observed as strong clustering
of quiescent galaxies around quiescent galaxies on angular scales $\theta\le
20$ arcsec, corresponding to a proper(comoving) scale of 168 (502) kpc at
$z=2$. The effect is observed only for quiescent galaxies around other
quiescent galaxies; the probability to find star-forming galaxies around
quiescent or around star-forming ones is consistent with the clustering
strength of galaxies of the same mass and at the same redshift, as observed in
dedicated studies of galaxy clustering. The effect is mass dependent in the
sense that the quenching probability is stronger for galaxies of smaller mass
($\rm{M_*<10^{10} M_{\sun}}$) than for more massive ones, i.e. it follows the
opposite trend with mass relative to gravitational galaxy clustering. The
spatial scale where the effect is observed suggests these environments are
massive halos, in which case the observed effect would likely be satellite
quenching. The effect is also redshift dependent in that the clustering
strength of quiescent galaxies around other quiescent galaxies at
$\bar{z}=1.6$ is $\approx 1.7\times$ larger than that of the galaxies with the
same stellar mass at $\bar{z}=2.6$. This redshift dependence allows for a
crude estimate of the time scale of environmental quenching of low--mass
galaxies, which is in the range $1.5\sim 4$ Gyr, in broad agreement with other
estimates and with our ideas on satellite quenching.
\end{abstract} 

\section{introduction} \label{sec:intro}
Observations both in the local universe and at high redshift have clearly shown that galaxies are characterized by a distinct bimodality of star--formation and dynamical properties and that is reflected in a corresponding bimodality of colors, morphology types, specific star formation rates \citep{Baldry2004, Baldry2006, Wyder2007, Blanton2009, Bamford2009}. Particular attention has been devoted to the physics of quenching, which refers to the sets of processes that shut down the star formation activity inside galaxies and drive the transformation of galaxies from one type of the bimodality to the other, i.e.  from a star-forming galaxy to a quiescent one. These processes remain observationally unconstrained.

While the detailed physical mechanisms of quenching are unclear, phenomenologically two broad categories of quenching mechanisms have been identified -- ``mass quenching'' and ``environmental quenching'' \citep{Peng2010, Schawinski2014}. Mass quenching generically refers to processes internal to a galaxy that depend on (or correlate with) the mass of the galaxy, like AGN and stellar feedback
\citep{Fabian2012, Hopkins2012}, morphological quenching \citep{Martig2009} or halo mass shock heating \citep{Dekel2006}. For example, the strong correlation between the presence of a massive bulge and the probability the galaxy is quenched \citep{Drory2007} has been interpreted as evidence that the central AGN may affect quenching \citet{Franx2008, Cheung2012, Barro2015}. \citet{Whitaker2017} reported a tight correlation between the central stellar surface density and the star formation activity, namely the fact that as galaxies quench they also develop a central structure characterized by high stellar mass density. This would imply a common mechanism (or mechanisms) controlling both the growth of the central regions of galaxies and the cessation of the their global star formation activity.

Unlike mass quenching, environmental quenching is associated with the
external environment of a galaxy and it is considered to be an effective
quenching mechanism of galaxies in dense environments (e.g galaxy
  groups/clusters). A number of specific mechanisms have been proposed
  for environmental quenching. For example, when a galaxy with a relatively
  small halo (satellite) is accreted by a massive halo, its gas supply from
  accretion from the cosmic web can be cut off. This will lead to a gradual
  quenching in a long-timescale as the satellite exhausts its own gas and is
  usually known as ``gas strangulation'' \citep{Larson1980, vandenBosch2008,
    Peng2015}. If the external pressure by the surrounding medium, i.e. the
  inter-cluster medium (ICM) or inter-group medium, is high enough, ram
  pressure stripping may also be able to remove cold gas from the satellite in
  a relative short-timescale, resulting in a rapid quenching
  \citep{Gunn1972}. Apart from the above two mechanisms, a process called
  ``galaxy harassment'' is also proposed for environmental
  quenching\citep{Farouki1981,Moore1998}, which refers the interactions
  between the satellite with high-speed fly-bys. The cumulative effect of many
  high-speed encounters can also significantly change the morphology of the
  satellite.

%Three main mechanisms have been proposed for environmental quenching
%-- ram pressure stripping, gas strangulation and harassment
%\citep{1972ApJ...176....1G,1980ApJ...237..692L,2008MNRAS.387...79V,2015Natur.521..192P}.
%When a satellite galaxy, which has a relatively small halo, is accreted by a
%massive halo, its gas can be stripped,leading to the shutdown of star
%formation. This process causes a long term decline of star formation activity
%inside a galaxy and is known as gas strangulation. If the external pressure is
%high enough, ram pressure stripping will rapidly remove cold gas and quench star
%formation quickly. The satellite galaxy may experience interactions involving high-speed 
%fly-bys, a process called harassment. The cumulative effect of many high-speed encounters can 
%significantly affect the morphology of the satellite galaxy.

%In terms of harassment, it is a mechanism that combines
%effects of tidal stripping and high-speed encounters. Harassment can change
%morphologies of satellite galaxies.

The correlations between stellar mass, star-formation and environment
  observed in the local Universe \citep{Gomez2003, Kauffmann2004,
      Balogh2004, Hogg2004, Blanton2005} have also been found to persist out
  to at least z$\rm{\sim}$1 \citep{Cucciati2006, Cooper2007, Peng2010,
    Sobral2011}. If these trends are indicative of both mass quenching and
  environmental quenching processes operating independently, then these
  processes must have already been in place by z$\rm{\gtrsim}$1. In fact,
\citet{Guo2017} find likely evidence of environmental quenching at $0.5<z<1$
based on the spatial distribution of low--mass
($\rm{8.0<Log_{10}(M_*/M_\sun)<9.5}$) quiescent galaxies around massive
($\rm{Log_{10}(M_*/M_\sun)>10.5}$) neighbors. Also, work by \citet{Lin2012} on
the clustering properties of bright BzK-selected galaxies at $z\sim2$ finds
evidence that the strength of galaxies' spatial clustering depends on their
star--formation properties, both star formation rate (SFR) and
  specific star formation rate (SSFR), which they interpret as evidence that
the environment has probably started to play a role in quenching star
formation already at that epoch.

Measuring the comparative strength of spatial clustering of galaxies as a function of their star formation activity indeed offers a powerful tool to investigate the phenomenology of quenching in galaxies at high redshift (e.g. $z>1$) and the correlations between star--formation activity and the environment, when large and well characterized samples are available (e.g. \citet{Coil2017}).  Spectroscopic observations of high redshift galaxies is resource--intensive, however, and even 8-10m telescopes can only observe relative bright galaxies and with a strong bias against quiescent galaxies. In particular, for statistical studies of spatial clustering, we do not have big enough spectroscopic samples of quiescent galaxies at $z>1$ to do robust spatial distribution analysis. Angular clustering, or other diagnostics of the relative angular proximity of galaxies (e.g. \citet{Guo2017}), however, provides a robust alternative. In particular, the availability of large and deep multi-band photometric surveys from space, CANDELS \citep{Grogin2011, Koekemoer2011}, and the consequent improvement of photometric redshifts and spectral energy distribution (SED) fitting techniques, means that we can now probe the correlation between environmental effects and star--formation activity in the high redshift universe.

In this work, we study the environmental effects on quenching galaxies at high redshift, where we consider the environment as the volumes immediately around galaxies (e.g. $\rm{r<R_{vir}}$). We use H-band selected galaxies in GOODS fields \citep{Giavalisco2004} from CANDELS. We measure the small-scale angular correlation function for different types of galaxies and investigate possible evidence of environmental effects on star-formation activity at redshift $\bar{z}\approx 2$ (mean redshift of our sample). In this paper, we adopt a $\Lambda$CDM cosmology with the parameters: $\Omega_m = 0.3$, $\Omega_\Lambda=0.7$ and $\rm{h=H_0/(100kms^{-1}Mpc^{-1})=0.7}$.

\section{Method}\label{sec:method}

The goal of this study is to investigate how star-formation activity changes with the environment at $\bar{z}\approx2$ by means of a comparative analysis of the strength of the angular clustering of quiescent and star-forming galaxies. In Section \ref{sec:data}, we describe the data selection and their division to quiescent and star-forming samples. In Section \ref{sec:analysis}, we present how we measure the angular clustering of the samples.

\subsection{The Data and The Samples} \label{sec:data}
Our main sample consists of 9887 galaxies culled from both the regions of the
GOODS-S ($\rm{\approx 0.05\; deg^2}$) and GOODS-N fields ($\rm{\approx 0.05\;
  deg^2}$) that have been observed with {\it HST}/WFC3 as part of the deep
portion of the CANDELS program \citep{Grogin2011, Koekemoer2011}. In this
  work, we have taken advantage of the deep CANDELS multi--wavelength
  photometry available in the GOODS fields and the official CANDELS
  photometric redshift catalog (see \citet{Dahlen2013,Hsu2014}) in which the
  full pdf is used in the determination of photometric redshift obtained with
  the EAZY code \citep{Brammer2008} and the templates by \citet{Muzzin2013}. A
number of papers have presented measures of stellar mass in the CANDELS fields
\citep{Tomczak2014, Santini2015, Mobasher2015}.  Here we have used the
measures of stellar mass obtained by \citet{Lee2018}, which adopts an advanced
{\it Markov Chain Monte Carlo} (MCMC) SED fitting procedure that treats the
star--formation history (SFH) of the galaxies as a free ``parameter'' to
obtain robust estimates of the stellar mass, star--formation rate and
(luminosity weighted) mean stellar age. These measures of stellar mass are in
excellent agreement with the other works (see Figure 5 in \citet{Lee2018}). We
have selected our sample galaxies to be in the redshift range $1.2<z<4$ using
spectroscopic redshift ($\rm{\approx 6\%}$) whenever available or photometric
ones and for having stellar mass $\rm{M_*>10^9}$ M$\rm{_{\odot}}$. To secure
high--quality photometry, and hence high-quality photo-z and SED-fitting
stellar mass measures, we have also required the isophotal H-band
signal--to--noise ratio to be SNR$>10$.

\begin{figure}
\gridline{\fig{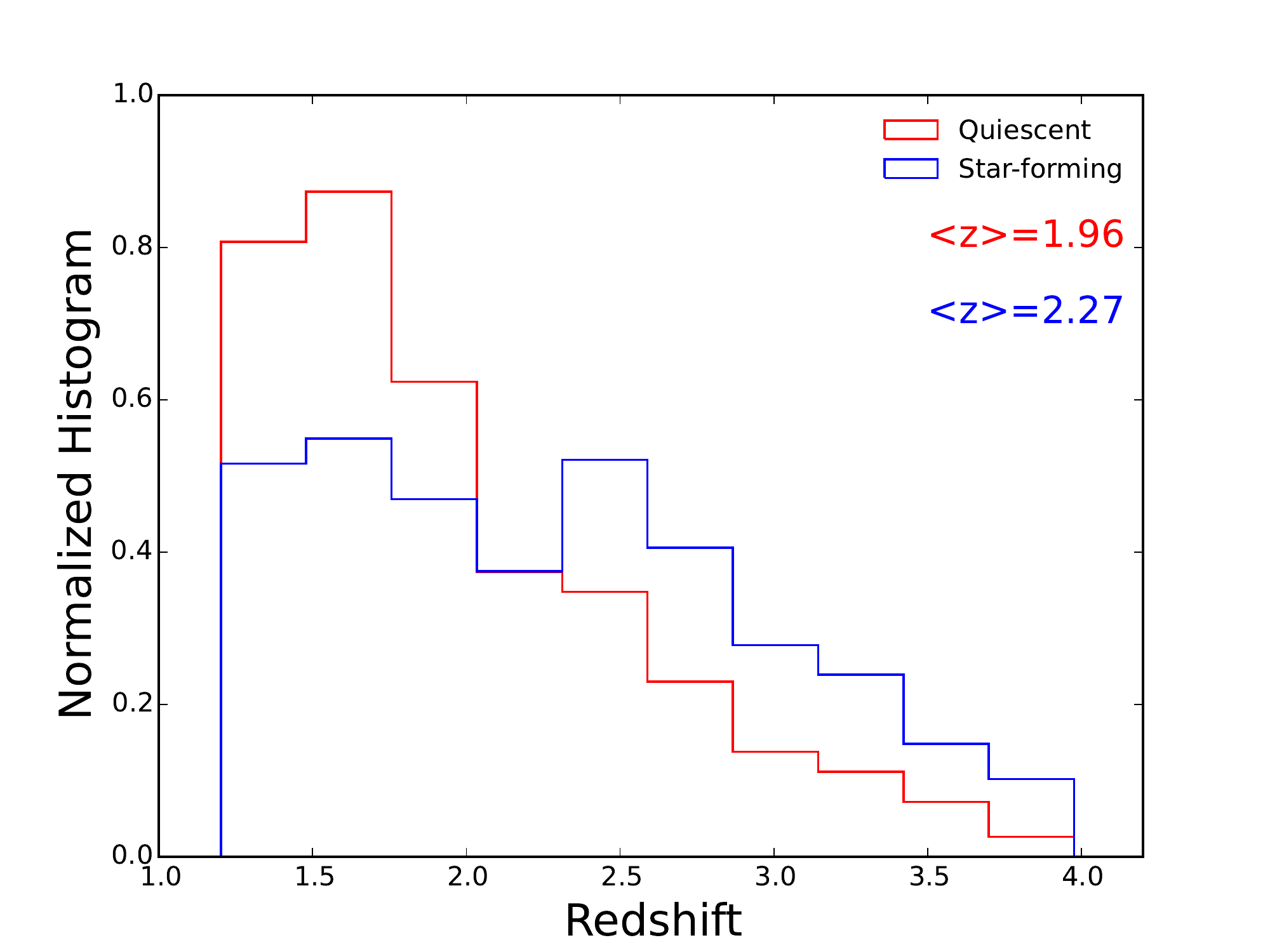}{0.23\textwidth}{(a)}
	\fig{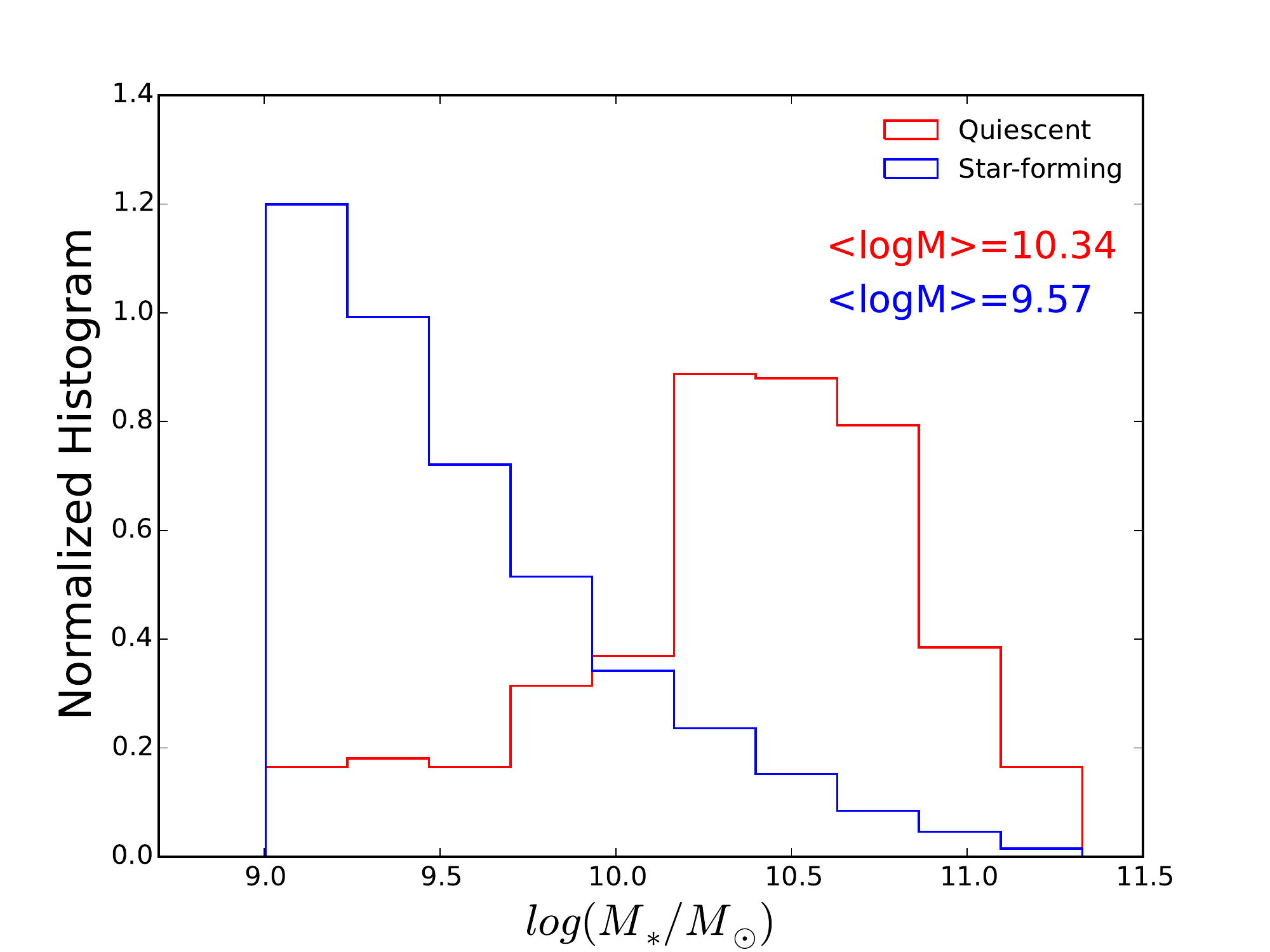}{0.23\textwidth}{(b)}
	}
\gridline{\fig{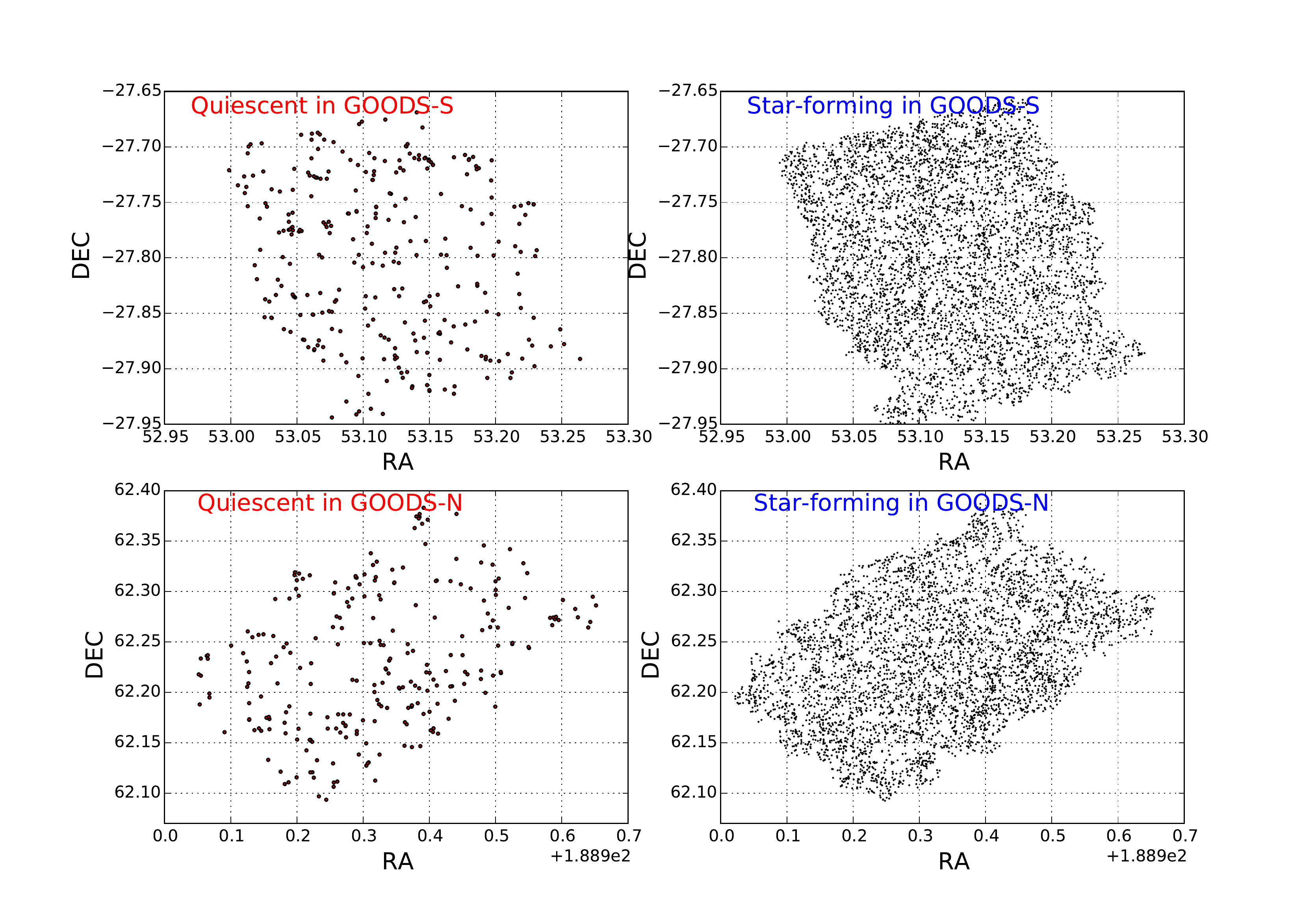}{0.5\textwidth}{(c)}
	}
\caption{(a) Redshift distributions for quiescent sample with mean redshift
  $\bar{z}=1.96$ and star-forming sample with $\bar{z}=2.27$, (b) Best-fit
  stellar mass ($\rm{M_*}$) distributions for quiescent sample with mean stellar
  mass $\rm{\overline{M_{*}}=10^{10.34}M_\odot}$ and star-forming sample with
  $\rm{\overline{M_*}=10^{9.57}M_\odot}$, (c) Angular distributions on the sky for quiescent and star-forming samples in GOODS-S and GOODS-N.} \label{fig:f1}
\end{figure}

\begin{figure}
\gridline{\fig{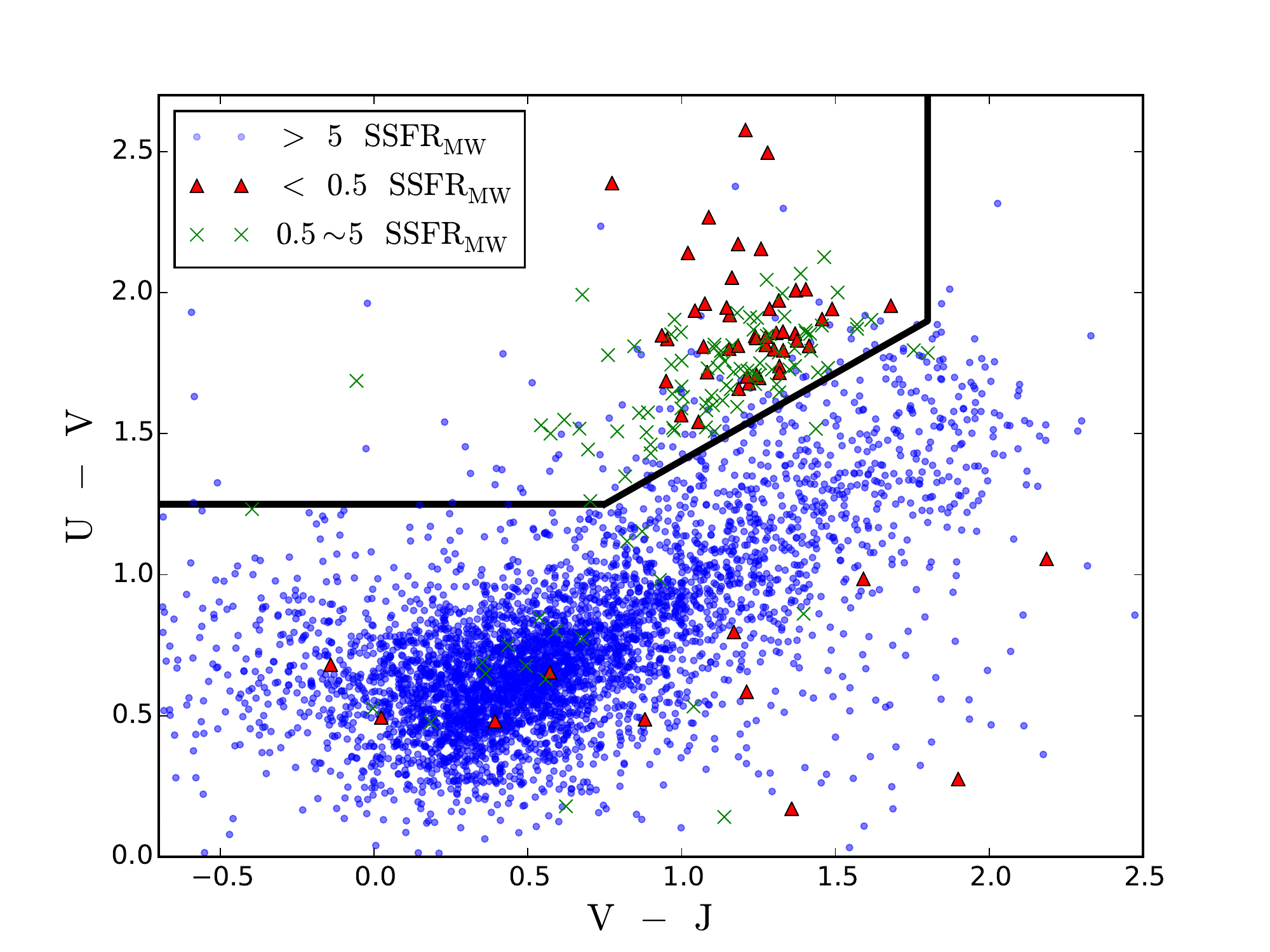}{0.5\textwidth}{}
	}
\caption{UVJ selection and comparison with SED fitting results. Three colors represent three SSFR bins, where we assume $\rm{SSFR_{MW}\sim 2/(8\times 10^{10}) yr^{-1}}$. Galaxies in the upper left region defined by the dash lines are defined as quiescent galaxies in this work.} \label{fig:UVJ}
\end{figure}

To classify our galaxies as quiescent or star-forming we have used the
UVJ-color selection method proposed by \citet{Williams2009}, for which we have
calculated the region of quiescent galaxies using spectral population
synthesis models from \citet{Bruzual2003}, illustrated in Figure
\ref{fig:UVJ}. We have also verified that the UVJ--selected samples of
quiescent and star--forming galaxies are in excellent agreement with an
analogous definition based on direct measures of SSFR, as discussed, for
example, in \citet{Lee2018}. In what follows, we define galaxies inside the
quiescent region of UVJ diagram as ``quiescent'' or ``quenched'' galaxies and
those outside as ``star--forming'' galaxies. The final quiescent sample
contains 294 galaxies in the GOODS-S field and 254 galaxies in the GOODS-N
field, while the star-forming sample includes 4977 galaxies in GOODS-S and
4362 galaxies in GOODS-N. Figure \ref{fig:f1} (a) and (b) show the (mostly
photometric) redshift and stellar mass distributions for the combined
star-forming and quiescent samples. Figure \ref{fig:f1} (c) shows the angular
distributions for the quiescent and star-forming samples in GOODS-S and
GOODS-N respectively. Our quiescent samples are not located in one or two
clusters, instead, they cover all across the two fields.

\begin{figure}
\gridline{\fig{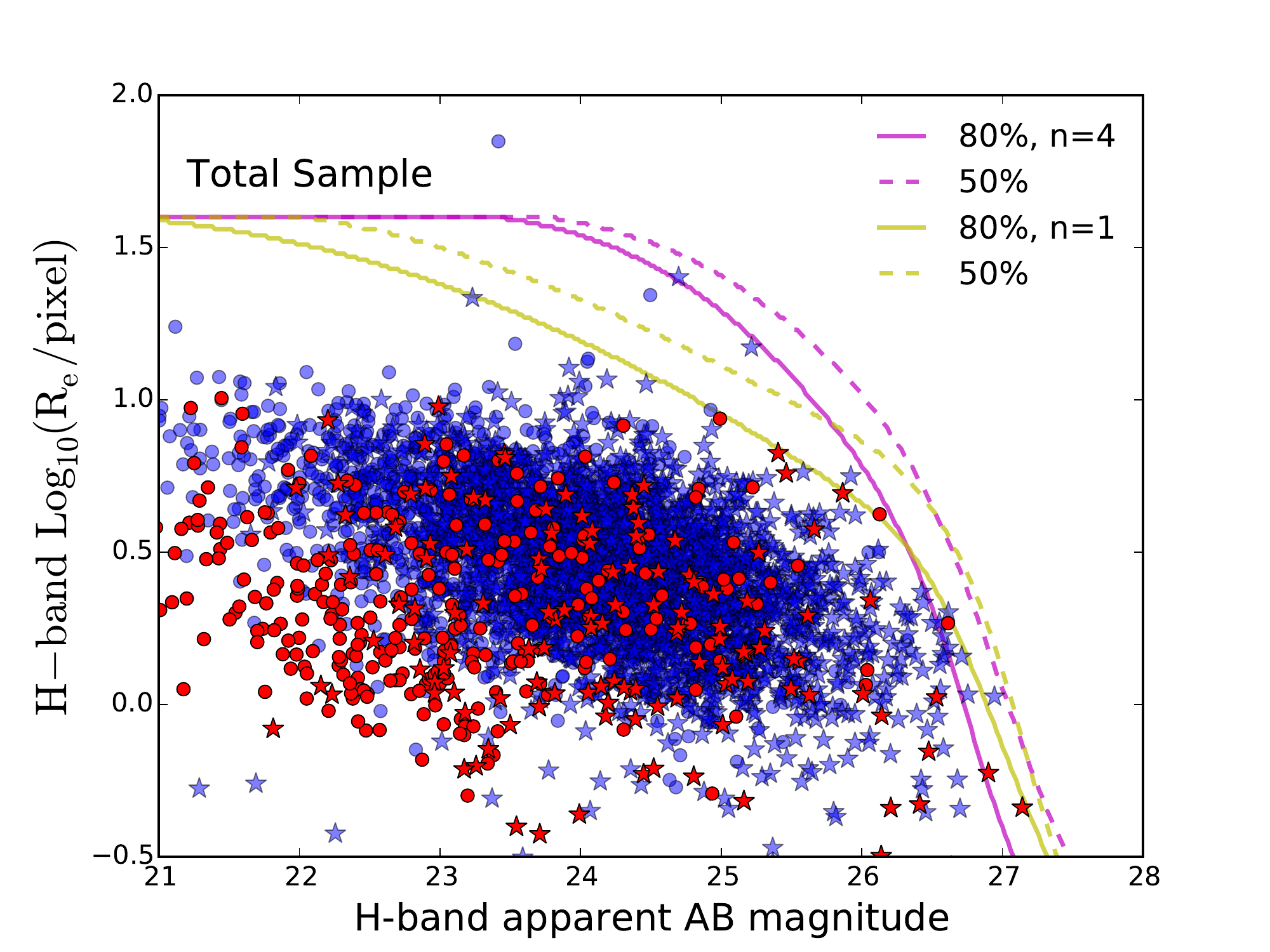}{0.5\textwidth}{}
	}
\caption{Completeness estimated from the simulations by \citet{Guo2013}. The light profile of fake sources is assumed to be an exponential disk ({\it Sersic index} n = 1, yellow) or a de Vaucouleurs profile (n = 4, magenta). In each case, the constant curves of 50\% (dashed) and 80\% (solid) completeness are plotted. Star-forming (blue) and quiescent (red) galaxies selected in this work are shown. The points and stars are for galaxies at $z<2$ and $z>2$ respectively.} \label{fig:complete}
\end{figure}

We have used the simulations by \citet{Guo2013} in GOODS to estimate the
completeness in our sample to be $\approx80\%$. As shown in Figure
\ref{fig:complete}, the majority of selected galaxies occupy the region where
sample completeness is $>80\%$. A few sources (mostly in the high redshift)
are in the region with completeness between 50\% and 80\%. As we will discuss
later, incompleteness does not substantially affect our conclusions because it
does not impact measures of angular clustering as long as there is no spatial
dependence on the probability for a galaxy of making into the samples or not,
which we do not observe. We will see that incompleteness only quantitatively
affects our measures of the quenched fraction,
i.e. the ratio of the number of quiescent galaxies to the total number of
galaxies. The magnitude of the effect, however, does not affect our
conclusions.

\subsection{Angular Clustering Amplitude Measurement} \label{sec:analysis}
The angular two-point correlation function $\omega(\theta)$ is defined as the excess probability, above that expected for a homogeneous ({\it Poisson}) distribution, of finding two galaxies with an angular separation $\theta$ within a solid angle $\delta \Omega$ \citep{Peebles1980} projected in the sky. In this work, we use the estimator of angular correlation function proposed by \citet{Landy1993}:
\begin{equation}
\omega(\theta) = \frac{DD(\theta)-2DR(\theta)+RR(\theta)}{RR(\theta)}
\end{equation}
where $DD(\theta)$ is the number of pairs of observed galaxies with angular separations in the range $(\theta, \theta+\delta \theta)$, $DR(\theta )$ is the number of cross-pairs between the observed galaxies and a randomly distributed sample and $RR(\theta)$ is the number of the randomly distributed pairs.

A random catalog of sources must be produced with the same sky coverage,
geometry and spatially-dependent detection incompleteness. We generate the
random samples by inserting 3000 randomly positioned sets of fake sources into
the noise map of GOODS-S and GOODS-N respectively. The inserted sources have
the same H-band magnitude distribution as what is observed for our galaxy
sample. We generate 20 of these simulations and select the random sources
($\rm{\approx 50000}$ in total) which are retrieved by SExtractor
\citep{Bertin1996} with 10 S/N. We calculate the angular correlation
  function of these retrieved random sources and have verified that on average
  the random sources in simulation are unclustered, i.e. $\omega(\theta)=0$.

We estimate the random errors on the two-point angular correlation function at each angular bin by two methods -- bootstrap resampling and spatial jackknife resampling. For bootstrap, we generate 100 resampling of the original sample, each containing N galaxies (including duplicates) randomly picked from the original N galaxies with replacement, i.e. a galaxy is retained in the stack even if it has already been picked. Then we estimated the error bar of bootstrap resampling by:
\begin{eqnarray}
\rm{\sigma^2_{bootstrap}\;=\;\frac{1}{N_{rs}}\sum_k^{N_{rs}}(\omega_k(\theta)-\bar{\omega}(\theta))^2}
\end{eqnarray}
where $\rm{N_{rs}=100}$, $\rm{\omega_k(\theta)}$ denotes the measurement of $\rm{\omega(\theta)}$ from the $\rm{k^{th}}$ resampling and $\rm{\bar{\omega}(\theta)}$ is the mean obtained from the 100 bootstrap resampling. For the spatial jackknife sampling, we quantify the error bars by binning GOODS-S and GOODS-N fields into 25 non-overlapping areas respectively and calculate the jackknife errors by
\begin{equation}
\rm{\sigma^2_{jackknife} =\frac{N-1}{N}\sum_{k=1}^{N}(\omega_k(\theta)-\bar{\omega}(\theta))^2}
\end{equation}
where N=25 areas, $\rm{\omega_k(\theta)}$ is calculated with the $\rm{k^{th}}$ area removed and $\rm{\bar{\omega}(\theta)}$ is the average values of $\rm{\omega_k(\theta)}$. As listed in Table 1, random errors on small angular scales ($\rm{<10}$ arcsec) estimated by these two methods are comparable. The jackknife errors in general are slightly smaller than those estimated by bootstrap resampling, so we conservatively adopt the bootstrap errors in this work. Estimating systematic errors of $\tt{\omega(\theta)}$ is more involved, since this needs to take into account the geometry and size of the observed field to model the strength of the Integral Constrains (IC) bias. As it will become clear later, our goal in this study is a comparison of the relative strength of the angular clustering of various sub--samples of galaxies extracted from the same main sample rather than the measure and fitting of the correlation function in each case. Since, to a
large extent, each measure of $\omega(\theta)$ is subject to the same IC bias, we have not included the correction because it will not affect the sense of the comparison of the relative strength of the clustering signal in our sub--samples.

\begin{deluxetable}{ccccc}
\tablecaption{Random errors estimated by bootstrap resampling and spatial jackknife resampling for the quiescent sample on small angular scales.}
\startdata
$\rm{Log_{10}(\theta/arcsec)}$&  & $\sigma_{bootstrap}$ & & $\sigma_{jackknife}$ \\
\hline
0.2 & &3.57 & &3.14 \\
0.4 & &1.74 & &2.30 \\
0.6 & &0.96 & &0.88 \\
0.8 & &0.51 & &0.44 \\
1.0 & &0.32 & &0.24 \\
\enddata
\end{deluxetable}

\begin{figure*}
\gridline{\fig{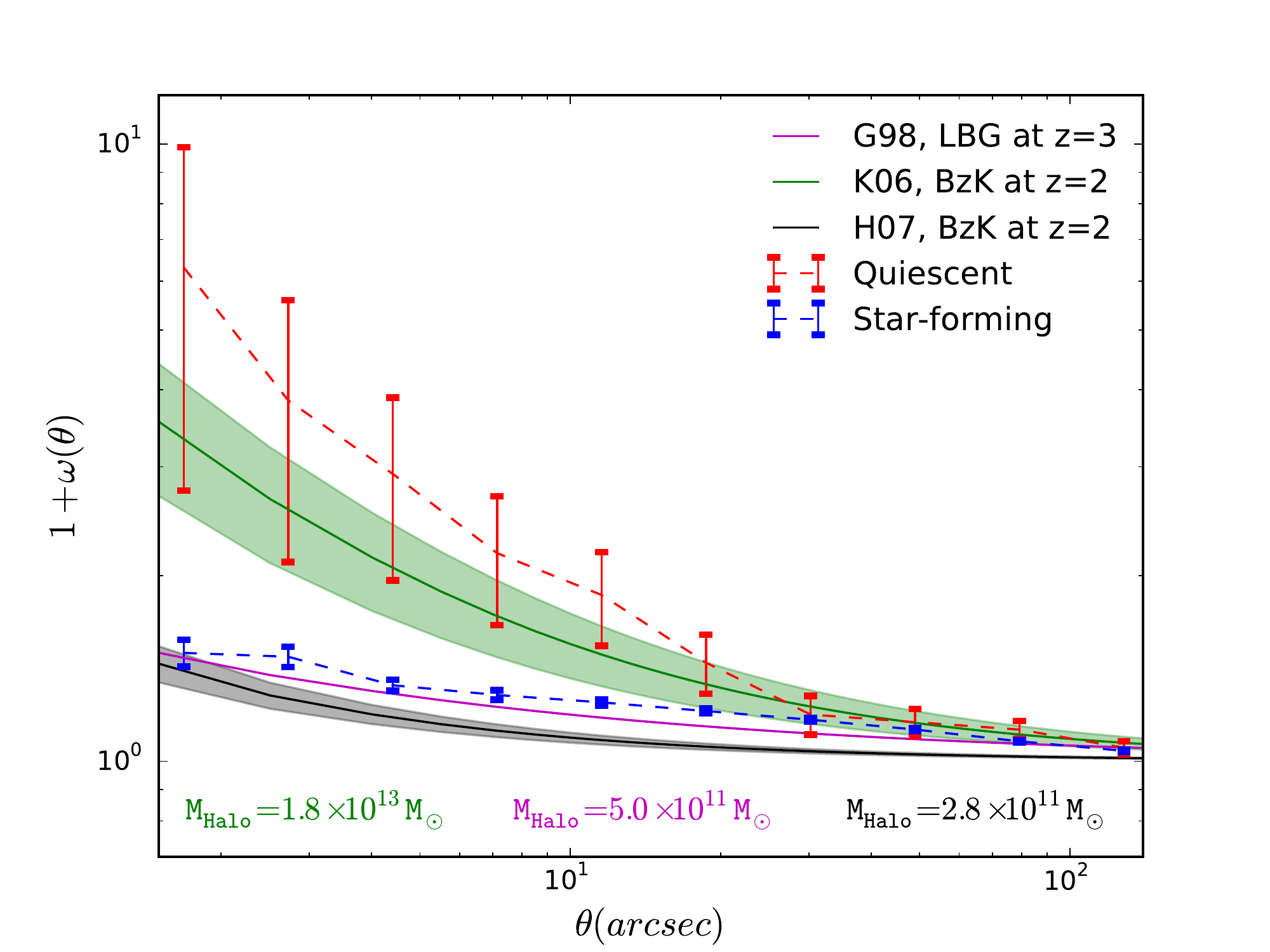}{0.9\textwidth}{}
	}
\caption{Results of $\tt{1+\omega(\theta)}$ for auto-correlation of quiescent galaxies (red) and star-forming galaxies (blue). Three power-law fitted angular correlation functions collected from literatures for other samples are also shown. The magenta solid line with $\rm{\omega(\theta) \sim 0.6\cdot\theta^{-0.5}}$ is for LBGs in G98. The best-fit power-law is obtained by \citet{Porciani2002} and halo mass of G98 is estimated as $\rm{M_{halo} = 5\times10^{11}M_\sun}$. The green solid line with $\rm{\omega(\theta) \sim3.46\cdot\theta^{-0.8}}$ (obtained by H07, see their Table 2) is for the subsample (sBzK with $\rm{K_{lim}=21.9Mag}$) of BzK-selected star-forming galaxies in K06. The halo mass of this subsample is estimated by H07 as $\rm{M_{halo} = 1.8\times 10^{13}M_\sun}$. The black solid line with fitted power-law of $\rm{\omega(\theta) \sim 0.58\cdot\theta^{-0.8}}$ is for H07 BzK-selected galaxies  and its halo mass is estimated as $\rm{M_{halo}= 2.8\times 10^{11}M_\sun}$. Green and black shaded regions indicate the uncertainty of the best-fit power-law for K06 and H07.}\label{fig:auto} 
\end{figure*}

Figure \ref{fig:auto} and \ref{fig:cross} show the measured
$\tt{1+\omega(\theta)}$ for the main sub--samples, namely the auto-correlation function of quiescent galaxies, auto-correlation function of star--forming galaxies and cross-correlation function of star--forming galaxies and quiescent galaxies. The figures (also Table 2) illustrate the main result of this study: the auto--correlation function of quiescent galaxies is much larger than that of star--forming galaxies, which has the same strength of the cross--correlation of quiescent with star--forming ones. Because the redshift distribution functions of all samples is similar (see Figure \ref{fig:f1}), differences in the angular clustering directly translates into similar differences in spatial clustering via the Limber transform (see \citet{Peebles1980}). In Figure \ref{fig:auto}, for comparison, we also show the power-law fitted angular correlation functions (already corrected for the IC) collected from literatures for other high redshift samples -- Lyman Break Galaxies (LBGs) at redshift 3 from \citet{Giavalisco1998} (G98), BzK color-selected galaxies at redshift 2 from \citet{Kong2006} (K06) and \citet{Hayashi2007} (H07).

The autocorrelation function of our sample of star-forming galaxies is quantitatively comparable to that of the low--mass sample of star--forming BzK galaxies from H07, which cover the same redshift range, and it is also similar to that of LBG at $z\sim 3$ (G98). Both types of star--forming galaxies are similar to our ones and should be hosted in dark--matter halos covering a similar mass range. The figure also shows the measured autocorrelation function of dark matter halos expected to host the high-mass star-forming BzK (K06) galaxies. In no case an excess at small angular scales of similar magnitude of the one of the quiescent autocorrelation function is observed.

Note that neglecting the IC correction is not likely to significantly affect the comparison of the relative clustering strength of the sub--sample we have considered. First and foremost, the IC correction relates to the large--scale behavior of the correlation function, where the effect of the finiteness of the samples are affected by the lack of knowledge of the number density of the parent population, while the small--scale clustering considered here is dominated by the structure of the halos. Secondly, when measured over sufficiently large volumes, the \citet{Landy1993} estimator of the angular correlation function that we have used here underestimates the true clustering due to the IC bias \citep{Hamilton1993}. This bias depends itself on the strength of the clustering of the galaxies being considered, and it is larger for more clustered galaxies (see \citet{Adelberger2005}; their Eqn. 13). Since we average together GOODS-N and GOODS-S fields and both the transverse and radial size of each field is much larger than the galaxy correlation length, neglecting the IC correction for the most strongly clustered sample results in underestimating its true strength more than it does for the more weakly clustered sample, which reinforces our conclusions.

Incompleteness, which affects our sample mostly in the high--redshift bin
  at the low end of the stellar mass distribution, does not significantly
  affect the results, and our conclusions, unless it is a function of the
  environment such that the incompleteness is higher in the field and lower in
  the dense environment, an occurrence for which there is no evidence. In
  fact, one would expect the opposite effect to happen because the background
  level and isophote confusion in a denser environment are higher than in the field, which would make the detection probability of galaxies in denser environments more incomplete than in the field.

\begin{deluxetable*}{cccccccc}
\tablecaption{Angular correlation functions in this work}
\startdata
$\theta/arcsec$ & $^{(a)}D/kpc$ & $^{(b)}\omega_Q$ & $^{(b)}\omega_{SF}$ & $^{(b)}\omega_{cross}$ &$^{(c)}\omega_Q/\omega_{SF}$&$^{(d)}\omega_{cross}/\omega_{SF}$\\
\hline
2.0 & 51.4 & 5.31$\pm$3.57 &  0.50$\pm$0.075 & 0.84$\pm$0.17 &10.66$\pm$7.34 & 1.68$\pm$0.43 \\
3.2 &  81.5 & 2.84$\pm$1.74 & 0.48$\pm$0.056 & 0.44$\pm$0.11& 5.96$\pm$3.72 & 0.92$\pm$0.26\\
5.1 & 129.2 & 1.92$\pm$0.96 & 0.33$\pm$0.027 & 0.33$\pm$0.08& 5.87$\pm$2.97 & 1.01$\pm$0.25\\
8.2 & 204.8 & 1.17$\pm$0.51 & 0.28$\pm$0.022 & 0.34$\pm$0.05& 4.19$\pm$1.86 & 1.22$\pm$0.21\\
12.9 & 324.6 & 0.86$\pm$0.32 & 0.25$\pm$0.015 & 0.26$\pm$0.04& 3.50$\pm$1.33 & 1.05$\pm$0.16\\
20.5 & 514.4 & 0.44$\pm$0.16 & 0.21$\pm$0.011 & 0.21$\pm$0.03& 2.16$\pm$0.78 & 1.02$\pm$0.17\\
32.5 & 815.3 & 0.19$\pm$0.09 & 0.17$\pm$0.009 & 0.19$\pm$0.03& 1.14$\pm$0.52 & 1.16$\pm$0.18\\
51.5 & 1292.1 & 0.16$\pm$0.06 & 0.13$\pm$0.006 & 0.17$\pm$0.03& 1.25$\pm$0.47 & 1.32$\pm$0.21\\
81.5 &  2047.9& 0.13$\pm$0.04 & 0.08$\pm$0.005 & 0.11$\pm$0.02& 1.63$\pm$0.49 & 1.41$\pm$0.30\\
129.2 & 3245.7 & 0.05$\pm$0.03 & 0.04$\pm$0.004 & 0.07$\pm$0.02& 1.32$\pm$0.66 & 1.72$\pm$0.48\\
\enddata
\tablenotetext{a}{Comoving scales at z = 2}
\tablenotetext{b}{Errors are estimated by bootstrap resampling}
\tablenotetext{c}{Ratio of auto-correlation of quiescent galaxies to that of star-formation galaxies}
\tablenotetext{d}{Ratio of cross-correlation of star-forming and quiescent galaxies to auto-correlation of star-formation galaxies}
\end{deluxetable*}

\section{Discussions} \label{sec:discussion}

\subsection{Evidence of environmental quenching}

As shown in Figure \ref{fig:auto} and listed as $\rm{\omega_{Q}/\omega_{SF}}$ in Table 2, the auto-correlation functions of quiescent and star-forming galaxies are similar at large scales ($\rm{\gtrsim}$ 30 arcsec, corresponding to $\rm{\gtrsim}$ 1 Mpc comoving scale). The excess clustering strength of the quiescent sample is mainly observed within $\rm{\lesssim}$ 20 arcsec. If we assume an average redshift $\bar{z}=2$ for the quiescent sample, this angular scale corresponds to a spatial proper (comoving) scale of $\rm{r_q\sim}$ 168 (502) kpc. The small spatial scale seems to suggest that the clustering signal originates inside massive halos, presumably the progenitors of large clusters, in which ram pressure stripping or gas strangulation should be expected to remove cold gas from infalling low--mass galaxies and terminate their star formation activity. However, since it is not possible for us to distinguish between low--mass galaxies that are satellites residing in massive halos or low--mass galaxies that are centrals of low--mass halos, the observed small--scale clustering actually depends on the relative proportions of satellites and centrals of the same mass. The observed excess clustering of quiescent galaxies therefore could be due to either the mass dependence of halo bias, increased satellite fraction of quiescent around other quiescent (caused by environmental effects) or both.

\begin{figure*}
\gridline{\fig{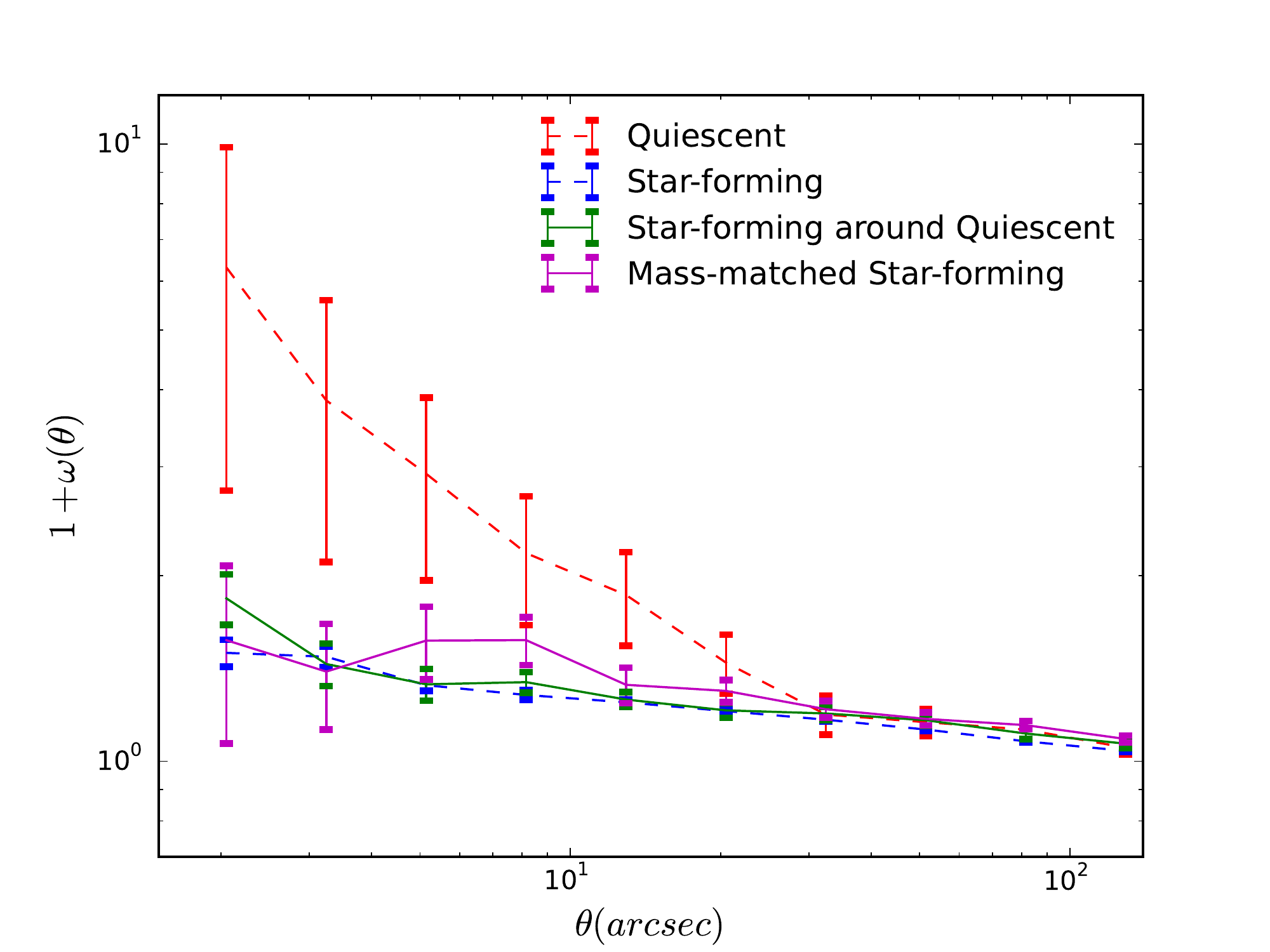}{0.9\textwidth}{}
	}
\caption{Results of $\tt{1+\omega(\theta)}$ for samples of star-forming galaxies that are mass-matched to the quiescent sample (magenta), star-forming galaxies around quiescent galaxies (green). For comparison, the results for quiescent sample and star-forming sample are also shown in the plot.} \label{fig:cross}
\end{figure*}

Qualitatively, on the same angular scales, because the function
$\omega(\theta)$ that we have measured is much larger than that of the much
more massive BzK star-forming galaxies at the same redshift selected by
K06 (compare the stellar mass distribution of our quiescent samples with
Figure 11(f) in K06), the large clustering strength that we observe at small
angular scales for the auto-correlation function of quiescent galaxies is
  unlikely to result from the hosting halos bias, which is an increasing
  function of halo mass and regulates the clustering strength of the general
  galaxy population. The fact that the two auto-correlation function at large
  scale ($\rm{\gtrsim}$ 1 Mpc comoving scale) are similar also indicates the
  similar ``two-halo'' term of quiescent and star-forming galaxies. Moreover,
  we measure the angular cross-correlation function of star-forming galaxies
  and quiescent galaxies (green solid line in Figure \ref{fig:cross}). This
  cross-correlation is, within the errors, the same as the auto-correlation
  function of star-forming galaxies, indicating that the halo structures of
  quiescent and star-forming ones are essentially the same. Thus, we interpret
  the observations as the the effects of environmental quenching on
  clustering, namely of the fact that within massive halos galaxies
  preferentially quench around other quiescent, more massive galaxies.

The stellar mass distributions of our quiescent sample and star-forming sample
are different (Figure \ref{fig:f1}(b)), with the former being more massive
than the latter. \citet{Coil2017} observed that at $z\approx 1.7$ the
clustering strength is a significantly stronger function of SSFR than it is of
the stellar mass, suggesting that the difference in stellar mass could only
play a minor role in the observed difference of clustering strength. We
further investigated whether stellar mass differences between our quiescent
and star-forming samples could account for the excess of angular clustering of
the quiescent galaxies in the following way. We measure the angular
auto-correlation function of mass-matched subsamples of star-forming galaxies
whose stellar mass distributions are the same as that of the quiescent sample
(magenta solid line in Figure \ref{fig:cross}). Compared with angular
auto-correlation function of star-forming galaxies, the angular clustering
increases slightly but the excess is still small compared with
auto-correlation of quiescent galaxies. The enhanced clustering signal is
consistent with the expected increase due to the larger typical host halo mass
of the mass-matched star-forming sample. This investigation indicates that the
excess clustering that we observe in the auto-correlation of quiescent
galaxies cannot be explained by the higher stellar mass of the sample alone.

\begin{figure*}
\gridline{\fig{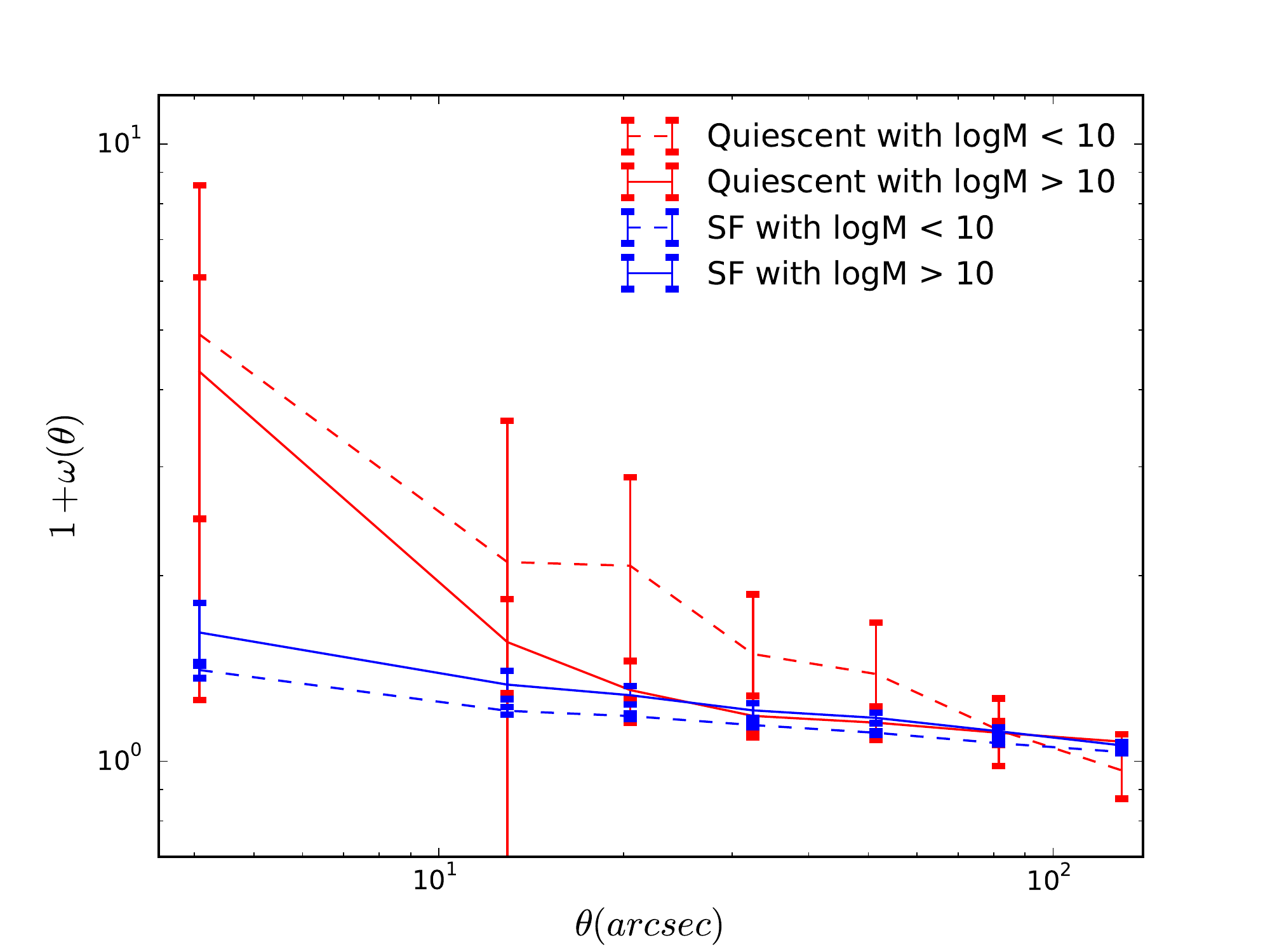}{0.9\textwidth}{}
    }
\caption{Results of $\rm{1+\omega(\theta)}$ for star forming galaxies (blue)
  and quiescent galaxies (red) with stellar mass $\rm{M_*>10^{10}M_{\sun}}$ and
  $\rm{M_*<10^{10}M_{\sun}}$. The low mass quiescent galaxies show stronger angular
  clustering than high mass quiescent galaxies. The low mass star forming
  galaxies show weaker angular clustering than high mass star forming
  galaxies. } \label{fig:f6}
\end{figure*}

As an additional test, we have divided our star-forming and quiescent samples
into two stellar mass bins, i.e. $\rm{M_* > 10^{10}M_{\sun}}$ and
$\rm{M_*<10^{10}M_{\sun}}$. We cannot distinguish central and satellite
galaxies in our sample. But, statistically, galaxies in low-mass bin should
include larger numbers of satellites (see e.g. \citet{Mandelbaum2006}).
Figure \ref{fig:f6} shows the angular clustering of star-forming and quiescent
galaxies in the two stellar mass bins. The low mass star-forming galaxies
have weaker clustering strength compared with high mass star-forming ones,
which is expected from normal gravitational clustering (i.e. the bias is an
increasing function of the dark halo stellar mass), for which more massive
galaxies have larger spatial clustering. But the figure shows that the 
clustering strength of quiescent galaxies in the two stellar mass bins folows
the opposite trend with stellar mass, with the lower mass bin having the
larger clustering strength by a factor $\approx 1.5$. 

We tested the significance of this difference using Monte Carlo simulations,
in which we took into account for the fact that the individual points of the
$\omega(\theta)$ function are not statistically independent but correlated
with the following procedure. The simulations test the null hypothesis that
the two observed correlation functions are actually two realizations of the
same parent population, i.e. both the high-- and low--mass samples have
the same angular clustering. We first run the simulations by treating each point of $\omega(\theta)$ as
independent and then we correct the results to account for the effect of
correlation between the points, which we estimate separately. At each angular
bin within 100 arcsec, we generate two set of simulated observations of
$\omega(\theta)$, one for the high-stellar mass sample and one for the
low-mas one, from two gaussian distributions with the same mean, assumed to be
equal to the observations of the high-mass $\omega(\theta)$, and with
variance equal to the error bar of each point. In this way, we automatically
take into account that the high-mass sample data points have smaller
uncertainty than the low mass ones. We then calculated the probability that
the $\omega(\theta)$ of the low-stellar mass sample is found to be smaller by
the observed amount at each angular separation point simultaneously. In $10^8$
realizations we found this probability to be $ 6.25\times 10^{-4}$ or $\approx
3.5\sigma$ in a gaussian statistics. To include the effects of the correlation between the points of
$\omega(\theta)$, which results in overestimating the significance of the
observed clustering difference of the two mass bins, we used the Monte Carlo
simulations by \citet{Giavalisco2001}, in which a large number of
realizations of galaxy samples is generated with specified intrinsic angular
clustering. The measure of $\omega(\theta)$ of each of these samples,
therefore, automatically includes the correlation between the points. By
repeating the same ``null hypothesis'' test, in one case using the full
``correlated'' simulated data set and in another case using two appropriate
averaged $\omega(\theta)$ functions (one for the high-stellar mass bin and
one for the lower one) as ``measures'' and treating its data points as
independent, we derived the correction function to be $\approx 2$. We
therefore conclude that the significance of the difference between the angular
clustering of the high-mass and low-mass bins is $\approx 1.8\sigma$. The fact that the
  strength of small scale clustering for the quiescent population is smaller
  in the more massive sample provides further evidence that stellar mass is
not the primary parameter that controls the clustering strength in this case,
and thus cannot be the reason of the much enhanced clustering of quiescent
galaxies compared to every other case. This also indicates that additional
factors, e.g environments, are required to explain the clustering excess which
is observed in Figure \ref{fig:auto}.

\subsection{Redshift dependence of environmental quenching}
\citet{Kawinwanichakij2016} measured the evolution of the quiescent fraction and quenching efficiency of satellites. For their sample, the satellite quenching is significant at $0.6<z<1.6$, while it is only weakly significant at lower or higher redshift. \citet{Hatfield2016} analyzed cross-correlation signal for their sample and conclude that at $z \sim 2$ environment is not a significant factor in determining quenching of star-forming galaxies. To understand the redshift dependence of environmental effects, it is crucial to constrain the timescale of environmental quenching from observations.  Direct measurements of the timescale of environmental quenching, especially at high redshift, however, remain uncertain because the they would require tracking or constraining the infall history of satellite galaxies, which is model dependent. According to recent studies, the quenching time for satellite galaxies is $4.4\pm 0.4$ Gyr at $z\sim 0.05$ (\citet{Wetzel2013}, based on data from the Sloan Digital Sky Survey (SDSS)); $1.05\pm 0.25$ Gyr at $z\sim 0.9$ (\citet{Mok2014}, who used the Group Environment Evolution Collaboration 2 (GEEC2)); $1.0\pm 0.25$ Gyr at $z\sim 1$ (\citet{Muzzin2014}, based on the Gemini CLuster Astrophysics Spectroscopic Survey (GCLASS)); and $2\sim5$ Gyr at $z\sim 1-2$ (\citet{Fossati2016}, 3D-HST). \citet{McGee2014} argued that the evolution of satellite quenching timescale could be caused by ``orbit-based" (e.g ram-pressure stripping) or ``outflow-based" mechanisms and the efficiency of these mechanisms could be different at high redshift. Therefore, estimating the timescale of environmental quenching is critical to constraining the mechanisms at play.

\begin{figure*}
\gridline{\fig{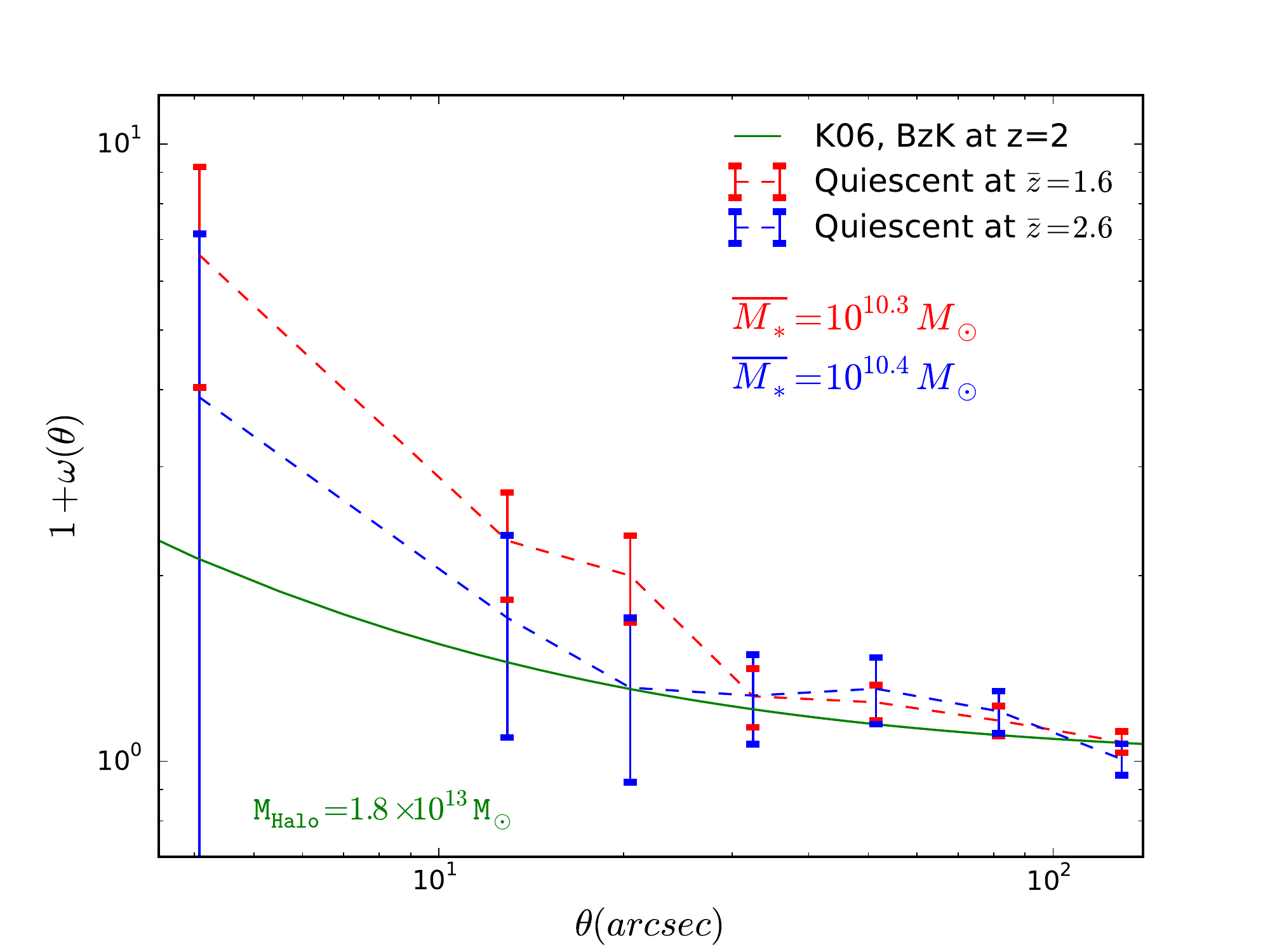}{0.9\textwidth}{}
    }
\caption{Results of $\rm{1+\omega(\theta)}$ for total quiescent galaxies at
  $z>2$ (blue) and $z<2$ (red). The measured $\rm{1+\omega(\theta)}$ for BzK
  galaxies with $\rm{M_{halo}=1.8\times 10^{13}M_\odot}$ from K06 is also shown in
  the plot. The angular clustering for quiescent galaxies at $z<2$ is much
  stronger than BzK galaxies in K06 while statistics for our quiescent
  galaxies at $z>2$ is not good enough for comparison.} \label{fig:corr_z}
\end{figure*}

\begin{figure}
\gridline{\fig{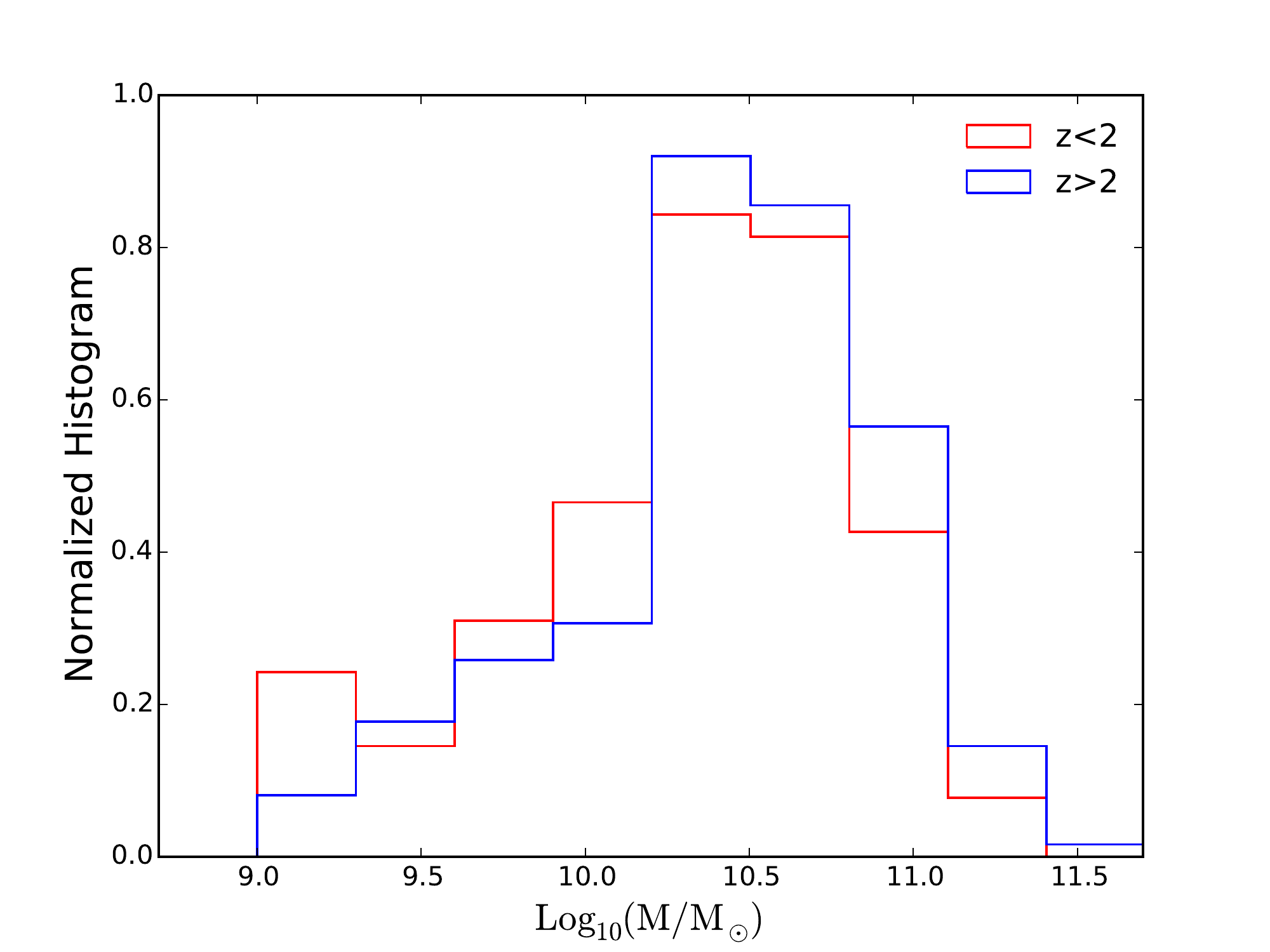}{0.5\textwidth}{}
    }
\caption{Normalized stellar mass distribution for the low--redshift (red) and high--redshift (blue) quiescent samples.}\label{fig:mass_dist}
\end{figure}

To provide a constraints to the timescale of environmental quenching, we have studied how the small-scale clustering of quenched galaxies has evolved with redshift. In particular, we have divided the quiescent sample into two redshift bins, one at $z<2$ (with mean redshift $\bar{z}=1.6$) and the other at $z>2$ (with mean redshift $\bar{z}=2.6$) and measured the angular clustering of both. As Figure \ref{fig:mass_dist} illustrates, these two samples have essentially  identical stellar mass distribution. If the redshift evolution of clustering were driven by the growth of structure, as is the case for the general mix of galaxies, the higher redshift sample should be more clustered because more biased relative to the average mass density distribution (e.g. see \citet{Adelberger2005,Lee2006,Lee2009,Tinker2010}). As shown in Figure \ref{fig:corr_z}, however, the clustering strength of quiescent galaxies around other quiescent galaxies at $\bar{z}=1.6$ is $\approx 1.7\times$ larger than that of the same galaxies with the same stellar mass ($\rm{M_*\approx10^{10.35}M_\sun}$) at $\bar{z}=2.6$, which is consistent with \citet{Kawinwanichakij2016} and \citet{Hatfield2016}. This is due to the appearance of low--mass quiescent galaxies, whose building up in the redshift range between $z\approx 2.6$ and $z\approx 1.6$ is illustrated in Figure \ref{fig:fq}. These low-mass galaxies are responsible for the observed strong small-scale angular clustering of quiescent galaxies, which is evidence of environmental quenching taking place around $z\approx 2$. Thus, a crude upper limit to the timescale of environmental quenching comes from the age of the universe at the mean redshift, $\bar z=1.6$, of our low--redshift quiescent sample, i.e. $\sim 4$ Gyr. Another approximate  estimate of the timescale over which significant environmental quenching of low--mass galaxies takes place comes from the difference of cosmic time between the average redshift, $\bar z=1.6$ and $\bar z=2.6$, of the two sub--samples provides, which is $\approx 1.5$ Gyr,  consistent with estimates from other groups, as reported earlier.

\subsection{Quenched fraction}
\begin{figure}
\gridline{\fig{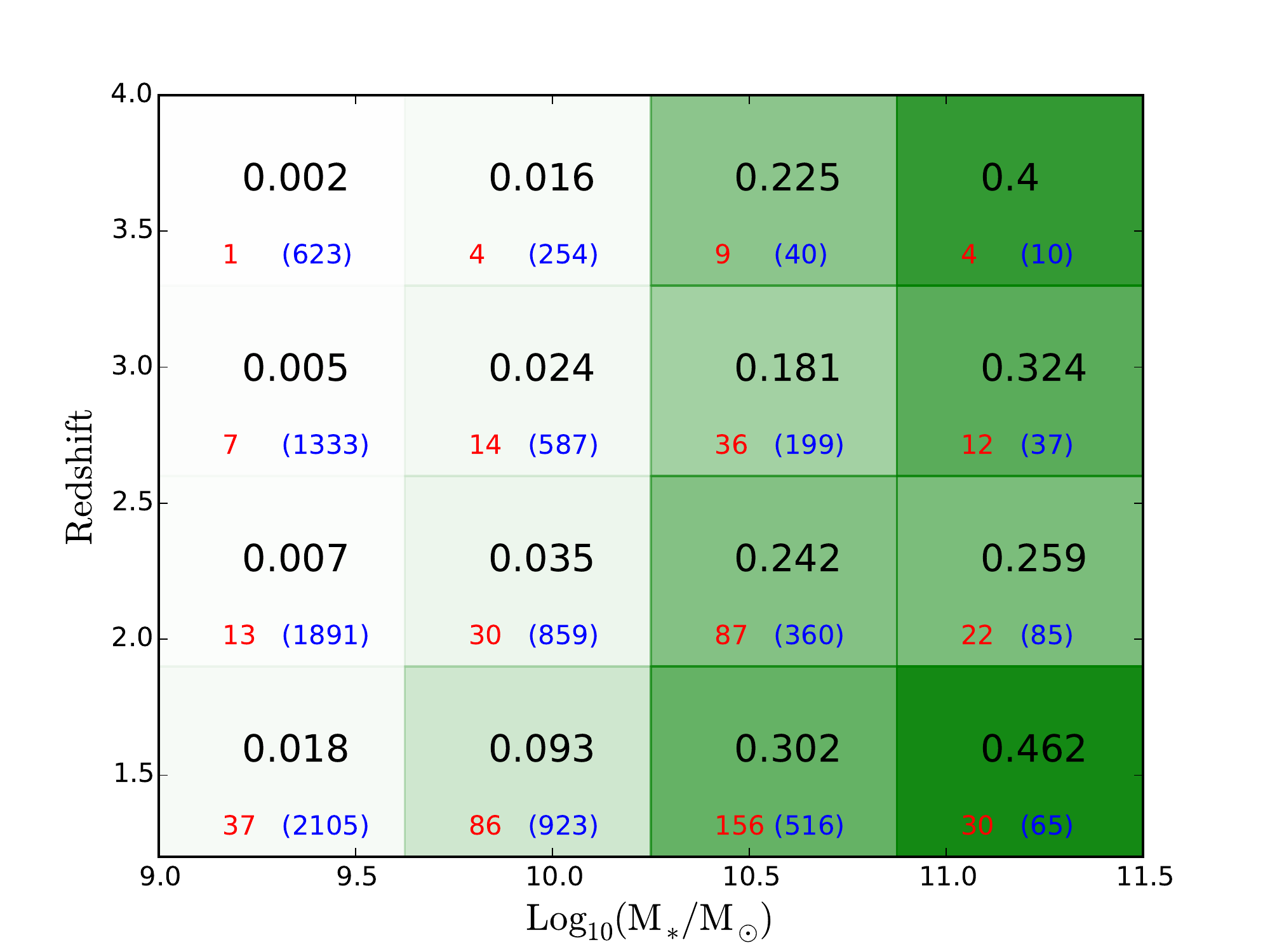}{0.5\textwidth}{}
    }
\caption{Quenched fraction as a function of stellar mass and redshift. In each bin, the number in black is the quenched fraction, the number in red is the number of quiescent galaxies and the number in blue is the total number of galaxies.} \label{fig:fq}
 \end{figure}

More insight into our angular clustering analysis can be gained from looking at the quenched fraction, i.e. the fraction of quiescent galaxies as a function of stellar mass and redshift, which is shown in Figure \ref{fig:fq}. Although our samples are quite complete (Figure \ref{fig:complete}), the effect of the relative incompleteness of star--forming and quiescent galaxies mush be tested first, since if the fractions of missed star-forming and quiescent galaxies differ this causes a systematic error in the shape of the quenched fraction. To check the effects of incompleteness, in particular to simulate the effects of missing fainter galaxies, we have measured the quenched fraction for the whole sample and also for two additional sub-samples obtained by selecting only galaxies which occupy the regions (with sample completeness $>80\%$) under the solid magenta and yellow curves defined in Figure \ref{fig:complete}. The quenched fraction for the full sample and for the two sub-samples are almost identical. The relative incompleteness therefore will not affect the measured quenched fraction much for our samples.

As Figure \ref{fig:fq} shows, the quenched fraction monotonically increases
with stellar mass at a fixed redshift, which is interpreted as the primary
evidence that there is key quenching mechanism which correlates with the
stellar mass, namely mass quenching \citep{Peng2010,Birrer2014}. For a fixed stellar mass, there is weak evidence that the quenched
fraction of high--mass galaxies ($\rm{Log_{10}(M_*/M_\odot)\gtrsim10}$)
evolves with redshift. This is not the case for low--mass galaxies
($\rm{Log_{10}(M_*/M_\odot)\lesssim10}$), however, whose quenched fraction
shows clear redshift dependence in the sense that the quenched fraction
increases as the redshift decreases. This is evidence that some other
mechanism, which is not mass quenching and which is significantly effective in
quenching low--mass galaxies but does not seem to effect high-mass ones, comes
into play as the Universe evolves.

Recall what has been extendedly discussed in previous sections, the excess clustering of quiescent galaxies on small-scale indicates that there is a quenching mechanism depending on proximity to other quiescent galaxies (i.e. environment). This environmental mechanism seems to be more profound in low--mass galaxies revealed as the inverse stellar mass dependence on auto-correlation of quiescent galaxies. Therefore, the increasing quenched fraction at low redshift bins is consistent with and very likely to be the result of environmental quenching.

\section{Conclusions and summary} \label{sec:summary}
The key observational result of this study is that the angular autocorrelation
function of quiescent galaxies on angular scales$\rm{\theta\le}$ 20 arcsec,
which corresponds to a spatial proper (comoving) scale $\rm{\approx}$
168 (502) kpc at $\bar{z}=2$ (the mean redshift of our sample), is much
stronger than that of the general population of galaxies of the same stellar
mass, i.e. selected regardless of the specific star formation rate. It is also
much stronger than that of galaxies hosting dark matter halos an order of
magnitude more massive. In other words, at redshift z $\rm{\approx}$ 2
quiescent galaxies cluster around other quiescent galaxies much more strongly
than the general galaxy population of the same stellar mass at the same
redshift and even $\rm{\approx}$ 2 $\rm{\times}$ more than galaxies hosting
more massive halos. Our measures are in qualitative agreement with the
  measures of the spatial transverse correlation function by \citet{Coil2017}
  in the sense that the clustering strength strongly depends on star formation
  activity of the samples, with galaxies of smaller star formation rate
having stronger clustering strenght.

While the strength of galaxy clustering generally increases with the
  stellar mass of galaxies because more massive galaxies are hosted in more
  massive dark matter halos and the bias of the halos is an increasing
  function of their mass, the opposite is seen for the quiescent galaxies in
  our sample, i.e. at small angular scales ($\theta\le 20$ arcsec) the
  clustering strength of auto-correlation of quiescent galaxies is stronger
  for lower mass ones. This inverse dependence on the stellar mass implies
  that the mechanism that increases the bias of quiescent galaxies at small
  scales must be related with the way these galaxies have quenched their star
  formation. The spatial scale of the observed excess clustering of the
  quiescent galaxies suggests that the environment of these galaxies are very
  massive halos, in which specific mechanisms, such as ram pressure, tidal  
  stripping or other causes of gas starvation and strangulation (our studies
  places no constraints on the specifics of such mechanisms) have
  shut down their star-formation activity. We therefore interpret these as
  the evidence of the manifestation of environmental quenching. We also
  measure the quenched fraction as a function of stellar mass and redshift,
  which provides evidence of the building-up of low--mass quiescent galaxies,
  in agreement with our conclusion that some mechanism 
  effective at quenching low--mass galaxies
  comes into play as the Universe evolves and consistent with our
  interpretation of environmental quenching. 

  The clustering strength of quiescent galaxies also varies with redshift in
  the sense that galaxies at z $\rm{<}$ 2 have higher clustering strength than
  those at z $\rm{>}$ 2. This is also consistent with environmental quenching
  because we expect the environmental quenching becoming more efficient as
  structures grow (notice that at fixed stellar mass, the clustering of the
  general population generally {\it increases} with increasing redshift
  because the galaxies are hosted in more massive halos). As we have
  discussed, this redshift dependence also enables us to put a crude estimate
  of the time scale of environmental quenching of low--mass galaxies,
  $\rm{\approx }$ 1.5 $\rm{\sim}$ 4 Gyr, which is consistent with results from
  other studies.

  Finally, our results also are in agreement with the similar study by
  \citet{Guo2017}, who use a slightly different statistical description of the
  angular separation of dwarf quiescent galaxies from the nearest massive
  ($\rm{M_*>10^{10.5}M_\sun}$) galaxy to reach essentially the same
  conclusions that this provides evidence of environmental quenching at z
  $\rm{\approx}$ 2.

\section{ACKNOWLEDGEMENTS}
We thank the anonymous referee for useful comments. The authors acknowledge the support for HST Programs GO-12060 and GO-12099 was provided by NASA through grants from the Space Telescope Science Institute, which is operated by the
Association of Universities for Research in Astronomy, Inc., under NASA
contract NAS5-26555.

%\bibliography{ji_2017_quenching}

\end{document}